\documentclass[useAMS,usenatbib]{mn2e}
\usepackage[utf8]{inputenc}
\usepackage{graphicx}
\usepackage{multicol}
\usepackage{color}
\usepackage{tabularx}
\usepackage{longtable}
\usepackage{hyperref}
\usepackage{graphics,epsfig,aas_macros,amsmath}

\newcommand{\asr}{{\it AstroSat~\/}}
\newcommand{\nus}{{\it NuSTAR~\/}}
\newcommand{\sxt}{{\it SXT~\/}}

\newcommand{\ig}{{ IGR J19294+1816~\/}}
\usepackage{xcolor}

\newcommand{\code}{\texttt}

\begin{document}

\title[\asr studies of IGR J19294+1816]{\asr detection of a mHz QPO and cyclotron line in IGR J19294+1816 during the 2019 outburst}

\author[G. Raman et al.]
{Gayathri Raman$^{1,2}$, Varun$^{3}$, Biswajit Paul$^4$ and Dipankar Bhattacharya$^5$  \\ 
$^1$ Indian Institute of Technology Bombay, Main Gate Rd, IIT Area, Powai, Mumbai, Maharashtra 400076, India\\
$^2$ Department of Astronomy and Astrophysics, The Pennsylvania State University, 525 Davey Lab, University Park, PA 16802, USA\\
$^3$ Aryabhatta Research Institute of Observational Sciences (ARIES), Manora Peak, Nainital-263001, Uttarakhand, India\\
$^4$ Raman Research Institute (RRI), Sadashivanagar, Bengaluru, Karnataka 560080, India\\
$^5$ Inter University Center for Astronomy and Astrophysics (IUCAA), Ganeshkhind, Pune, Maharashtra 411007, India}


 \maketitle

\begin{abstract}
We present results from timing and spectral analysis of the HMXB X-ray pulsar IGR J19294+1816 observed using \asr during its recent Type-I outburst in October, 2019. \asr observations sampled the outburst at the decline phase right after the outburst peak. We carried out timing analysis on the light curves obtained using the Large Area X-ray Proportional Counter (LAXPC) instrument on board \asr and measured a spin period of 12.485065$\pm$0.000015~s. The peak in the power density spectrum (PDS) corresponding to the spin period of 12.48~s also shows a broadened base. We also detected a Quasi Periodic Oscillation (QPO) feature at 0.032$\pm$0.002~Hz with an RMS fractional amplitude of $\sim$18\% in the PDS. We further carried out a joint spectral analysis using both the Soft X-ray Telescope (SXT) and the LAXPC instruments and detected a Cyclotron Resonant Scattering Feature (CRSF) at 42.7$\pm$0.9~keV and an Fe emission line at 6.4$\pm$0.1~keV. IGR J19294+1816, being an intermediate spin pulsar, has exhibited a plethora of spectral and timing features during its most recent 2019 outburst, adding it to the list of transients that exhibit both a QPO as well as a CRSF.

\end{abstract}


\section{Introduction}

Accretion powered High Mass X-ray Binary (HMXB) pulsars comprise of a highly magnetized neutron star (NS) accreting from a high mass companion. Most of these systems in the Milky Way galaxy are transient sources that usually remain in quiescence for long periods, and occasionally exhibit outbursts with durations ranging from several days to hundreds of days (see \citealt{Paul-Naik2011} for a review). Outbursts from these systems have been classified as Type-I and Type-II depending upon their peak luminosities and duration. Type-1 outbursts tend to be periodic with durations of a few days to tens of days and X-ray luminosities L$_{\rm X}\sim$10$^{35}$-10$^{37}$~ergs s$^{-1}$. They exhibit periodic enhanced episodes of accretion during the periastron passage of the NS. The Type-II outbursts, however, are not periodic, are much longer in duration, lasting for weeks to months and are characterized by luminosities $L_{X}${$\sim10^{37}-10^{38}$} ergs s$^{-1}$. These systems exhibit luminosity dependent timing and spectral features such as variations in pulse profiles, presence/absence of Quasi Periodic Oscillations (QPOs) in the power density spectra, variation in cyclotron line energy properties, etc. \citep{Paul-Naik2011}.

Accretion onto a NS is an efficient way of generating high energy photons. However, the maximum luminosity that can be reached by this process is limited by several factors. For the case of spherical accretion, the Eddington luminosity limits the maximum attainable luminosity. However, for a highly magnetized NS in an accreting binary system, 
the maximum luminosity attained, can greatly exceed the L$_{\rm Edd}$ via the formation of accretion columns \citep{Basko1976,Mushtukov2015}. From the current understanding of the theory of accretion, we know that emission region configuration strongly depends on the mass accretion rate \citep{Basko1976}. At lower X-ray luminosities, the accretion mounds on the NS polar caps majorly dominate the X-ray emission. The NS surface is heated up by the inflowing plasma via Coulomb collisions \citep{ZelShakura,Mushtukov2015}.
However, at higher mass accretion rates, the plasma is stopped at some distance above the NS surface by radiation-dominated shock and the radiation is observed to originate from an extended `accretion column' \citep{Basko1976,Becker2012,Mushtukov2015,Doroshenkov2017}.
A \textit{`critical'} luminosity has been defined in literature that serves as a boundary between these two regimes. The behavior of pulse profiles and even CRSF line energy properties have been shown to vary in these two domains \citep{Becker2012, Mushtukov2012}. Luminosity-related variations in the CRSF line energies and pulse profiles, help probe changes in the configuration of the accretion column, all of which can be best studied while accreting magnetized NS systems undergo episodic outbursts.

 Observations of X-ray pulsars have indicated that they have spin periods ranging from milli-seconds (for example, SAX J1808.4+3658, \citealt{Wijnands_klis1998}) all the way up to a few thousand seconds (for example, 4U2206+54, 2S 0114+65, see \citealt{Wang2020}). In fact, recent studies have indicated that the Be X-ray binary population comprises of two sub populations based on their spin and orbital periods and orbital eccentricities \citep{Knigge2011}: i) the short spin P$_{\rm spin} \sim$ 10~s,  P$_{\rm orb}\sim$40~d, and ii) P$_{\rm spin} \sim$ 200 s, P$_{\rm orb} \approx$ 100 d \citep{Priegen2020}. Such a bimodal distribution in the NS spin has been suggested to be the result of two distinct NS forming supernova pathways \citep{Knigge2011}. In this context, particularly interesting are the X-ray pulsars hosting NS with intermediate spin ($\sim$ several seconds to several tens of seconds) since they also occupy a unique position in the Corbet diagram of spin period versus orbital period. IGR J19294+1816 is one such poorly studied intermediate spin pulsar that underwent an outburst in October 2019, which we observed and studied using \asr instruments.
 
IGR J19294+1816, a hard X-ray transient, was first detected with the IBIS/ISGRI camera on board \textit{INTEGRAL} Gamma Ray Observatory while undergoing an outburst in 2009  \citep{Atel2}. Pulsations at 12.4~s were detected from this bright source using Swift observations \citep{Rodrig2009}; \citet{Strohmayer2009ATel} further confirmed the X-ray pulsar nature of this object using RXTE-PCA observations. \citet{Corbet-Krim2009} inferred an orbital period of 117.2~days using long term Swift-BAT flux modulations. NIR spectroscopy revealed that the system hosts a B1Ve optical companion star and the distance to the system was inferred to be around 11$\pm$1~kpc \citep{Tsygankov2019}. A cyclotron absorption feature was detected at $\sim$42~keV using \nus observations \citep{Tsygankov2019}.

A recent XMM-Newton observation detected enhanced X-ray activity during its periastron passage, indicating it to be a Type-I X-ray outburst \citep{Atel1}. In order to study the wide band spectral and timing characteristics of this source during its latest outburst, we carried out X-ray observations using the various \asr instruments. The paper is organized as follows. In Section 2, we present the details of the observations and data reduction methods for the LAXPC and SXT instruments. This is followed by timing and spectral analysis and results in Section 3. We then discuss the implications of our results and describe the broadband outburst properties of this pulsar system in Section 4. 

\section{Observations and Data Reduction}\label{sec:obs}

\asr observations of IGR 19294+1816 were carried out  using the SXT and LAXPC instruments on the 29th of October, 2019, following an ATel reporting re-brightening of the source in XMM-Newton observations \citep{Atel1}. The source was already in the declining phase of its outburst after its peak had passed. This observation was part of a Target of Opportunity (ToO) proposal (Obs ID T03 3272) comprising of \asr orbits 22082-22093 and a total exposure time of $\sim$62~ks. 
 
 The Indian satellite \asr carries five payloads providing simultaneous multi-wavelength coverage in the UV, soft and hard X-ray bands. The Large Area X-ray Proportional Counter (LAXPC) is a broad-band timing instrument that comprises of three co-aligned proportional counter detectors with a 1$^\circ \times$1$^\circ$ field of view. It achieves a timing resolution of 10~micro-second and is sensitive in the broad band energy range of 3.0--80.0~keV. LAXPC has proven to be extremely useful in carrying out CRSF line studies \citep{Varun2019new,Varun2019-new}. Details of the LAXPC calibration can be found in \citet{Antia2017,Yadav2017-new}. During this observation, the LXP30 detector was non functional and the LXP10 detector was corrected for gain changes rendering it unreliable for spectroscopic work. We have therefore utilised the LXP20 detector for spectroscopic analysis in this work. The level 1 LXP20 data files were reduced using the standard data reduction pipeline\footnote{http://astrosat-ssc.iucaa.in/?q=laxpcData} (Format A). The level 2 event file was generated using the standard procedure \code{laxpc$\_$make$\_$event}. We further generated the good time intervals that exclude the time intervals corresponding to Earth occultation and satellite passage through the South Atlantic Anomaly (SAA) using the task \code{laxpc$\_$make$\_$stdgti}. With the help of the \code{as1bary} tool, the new level 2 event file was then barycenter corrected from the satellite frame to the Solar System Barycenter using the orbital information and the source coordinates. The resultant event file was then used for further timing and spectral analysis.

\begin{figure*}
    \centering
    \includegraphics[scale=0.5,angle=-90,trim={0cm 0cm 0 0cm},clip]{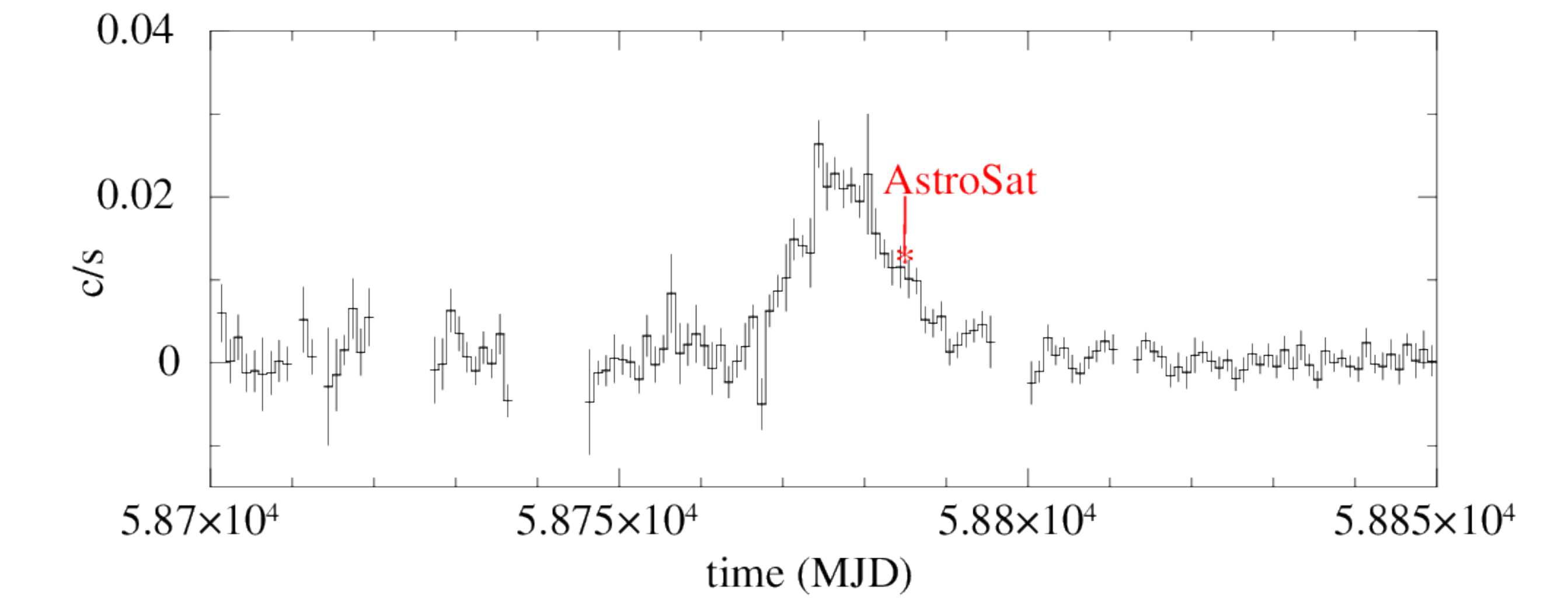}
    \caption{The 15--50~keV Swift BAT light curve shows the October 2019 outburst of IGR J19294+1816. A prompt \asr observation was undertaken and its location on the long term light curve is indicated in red.}
    \label{fig:batlc}
\end{figure*}

The Soft X-ray Telescope (SXT) is an imaging telescope operating in the 0.3--8.0~keV energy band \citep{Singh2016}. It has an X-ray optics unit consisting of 40 confocal cells made of gold-coated aluminium foils. The X-ray CCD consists of 600$\times$600 pixels. It images a circular region of the sky with a radius of $\sim20^{'}$. The on-axis Point Spread Function (PSF) has a full width at half maximum (FWHM) of $100^{\arcsec}$. It has a spectral resolution of $\sim$150~eV at 6~keV. During this observation, the \sxt was operated in the Photon Counting (PC) mode with a time resolution of 2.4~s. The LEVEL 1 PC mode data were processed using the \sxt software pipeline version 1.4b and the SXT spectral redistribution matrices in \code{CALDB (v20160510)}.  Orbit-wise event files were merged using the Julia software-based \sxt merger tool called \code{sxtevtmergerjl}. We selected all the events inside a circular region of radius $15^{'}$ centered on the source coordinates. The image, light curve and spectra were extracted using the \code{Xselect} tool. 

\section{Results}

\subsection{Timing}
\begin{figure}
    \centering
    \includegraphics[scale=0.33,angle=-90,trim={0cm 0cm 0cm 2cm},clip]{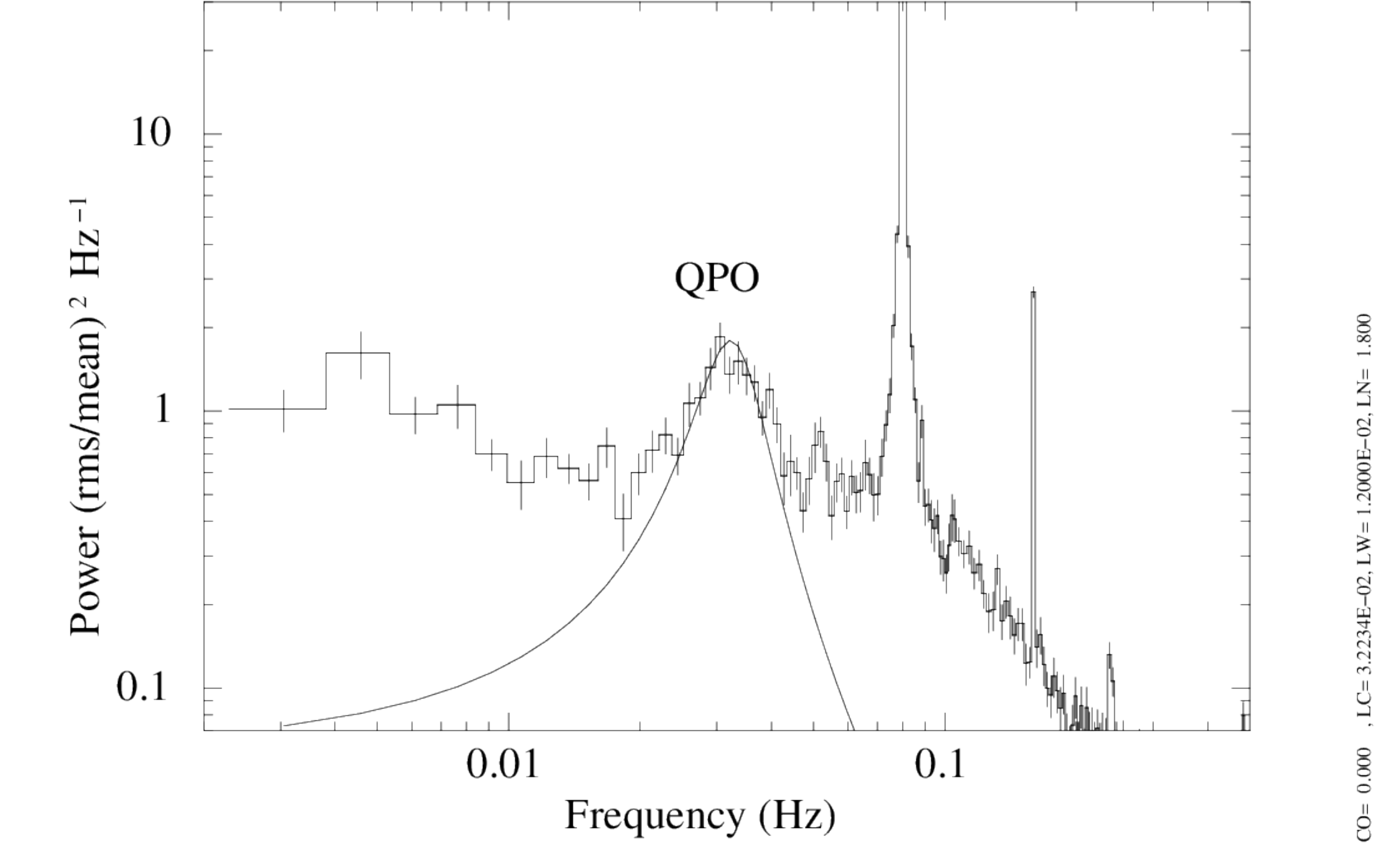}
    
    \includegraphics[scale=0.37,angle=0]{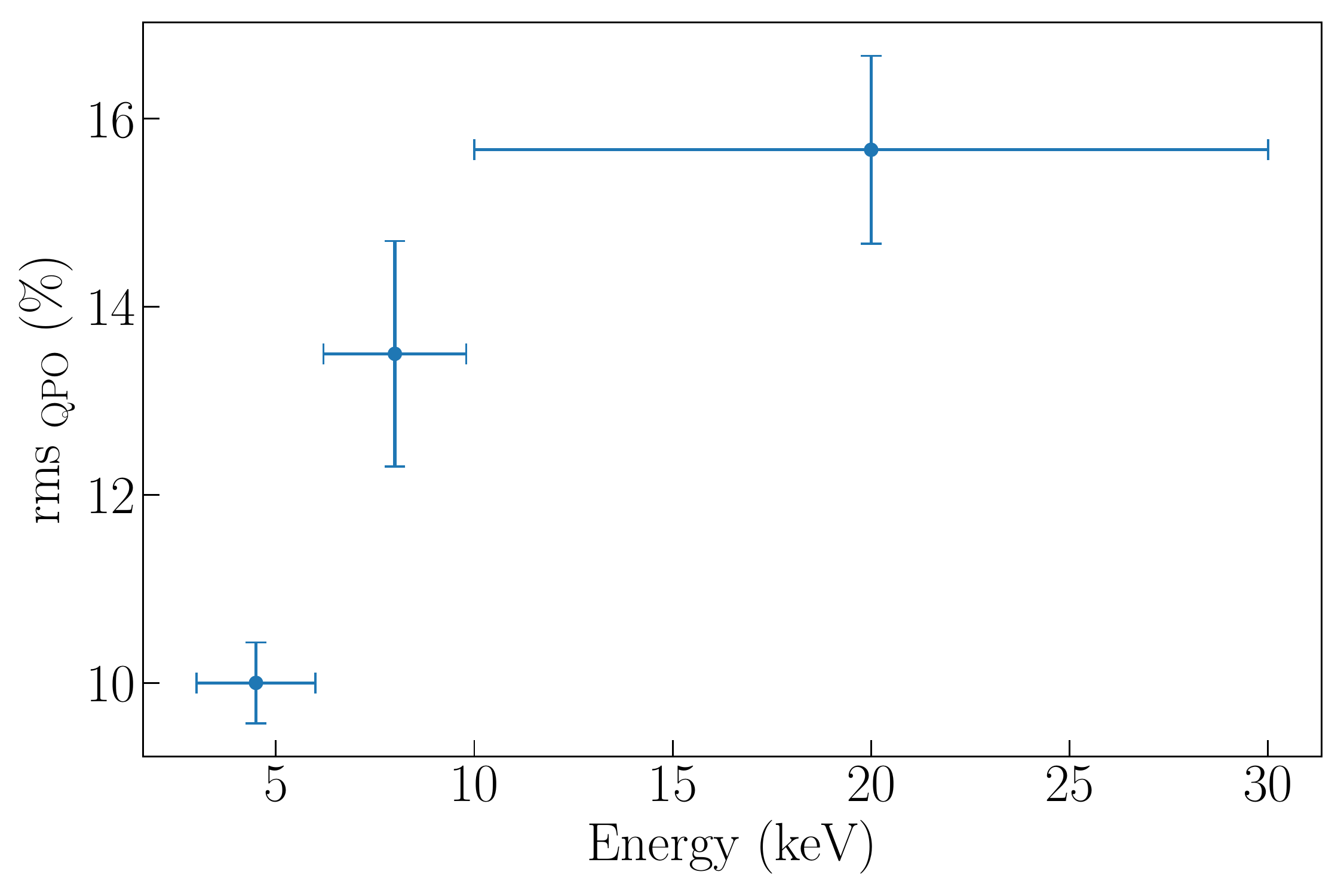}
    
    \caption{The white noise subtracted RMS normalised PDS of the light curve showing the pulse period at $\sim$0.08~Hz and the QPO feature at 0.032~Hz (top panel). The QPO is modeled using a Lorenztian function, shown here as a solid black line. The bottom panel shows the variation of the QPO rms fractional amplitude as a function of energy. The QPO is not detected beyond 30~keV.}
    \label{fig:pds}
\end{figure}
  
 We obtained a long term Swift BAT light curve from the Swift-BAT hard X-ray transient Monitor page\footnote{https://swift.gsfc.nasa.gov/results/transients/index.html} to indicate the segment of the outburst which was sampled by \textit{AstroSat}. The Swift-BAT light curve in the 15--50~keV energy band, with the \asr observation marked on it, shows a marked increase in the count rate during the October 2019 outburst (see Figure \ref{fig:batlc}).

We extracted the time series from LAXPC unit LXP20 with a 5~ms binning in the 3.0--80.0~keV energy band using the task \code{laxpc$\_$make$\_$lightcurve}. The background subtracted, barycenter corrected light curve showed a steady emission during the observations with an average count rate of 96~counts~s$^{-1}$.

We created the Power Density Spectrum (PDS) using the FTOOLs task \code{powspec} from the LXP20 light curve of the full observation. The PDS was normalised such that the integral gives the squared RMS fractional variability. Interestingly, we detect signatures of a QPO like bump at $\sim$0.032~Hz along with spin frequency peak at 0.08~Hz and its two harmonics (at $\sim$0.16~Hz and $\sim$0.25~Hz) in the 3.0--80.0~keV PDS as shown in Figure \ref{fig:pds}. The spin peak feature shows a peculiar broadening at its base. To infer the properties of the QPO, the PDS in the frequency range around the centroid of the QPO (0.01-0.07~Hz) was fitted with a model consisting of a constant to describe the continuum along with a Lorentzian component. The QPO centroid frequency was determined to be 0.032$\pm$0.002~Hz with a width of 0.012$\pm$0.006~Hz. The Quality-factor for this QPO feature is calculated to be (Q=$\nu_{\rm 0}/\sigma$) $\sim$2.7. The white noise subtracted RMS value of the QPO in the 3.0--80.0~keV energy band is found to be 13.6\%$\pm$0.5\%. We further calculated the QPO RMS fractional variability in different energy bands and found that it increased as a function of energy (Figure \ref{fig:pds}, bottom panel). The QPO is detected in the 3 energy bands 3--6~keV, 6--10~keV, 10--30~keV with an RMS of 10.2\%$\pm$0.4\%, 13.5\%$\pm$1.2\%, and 15.7\%$\pm$1.0\%, respectively. The QPO feature is not detectable in the LAXPC power spectra beyond 30~keV. 
 
\begin{figure}
    \centering

    \includegraphics[scale=0.5,angle=-90,trim={0cm 0cm 0cm 0cm},clip]{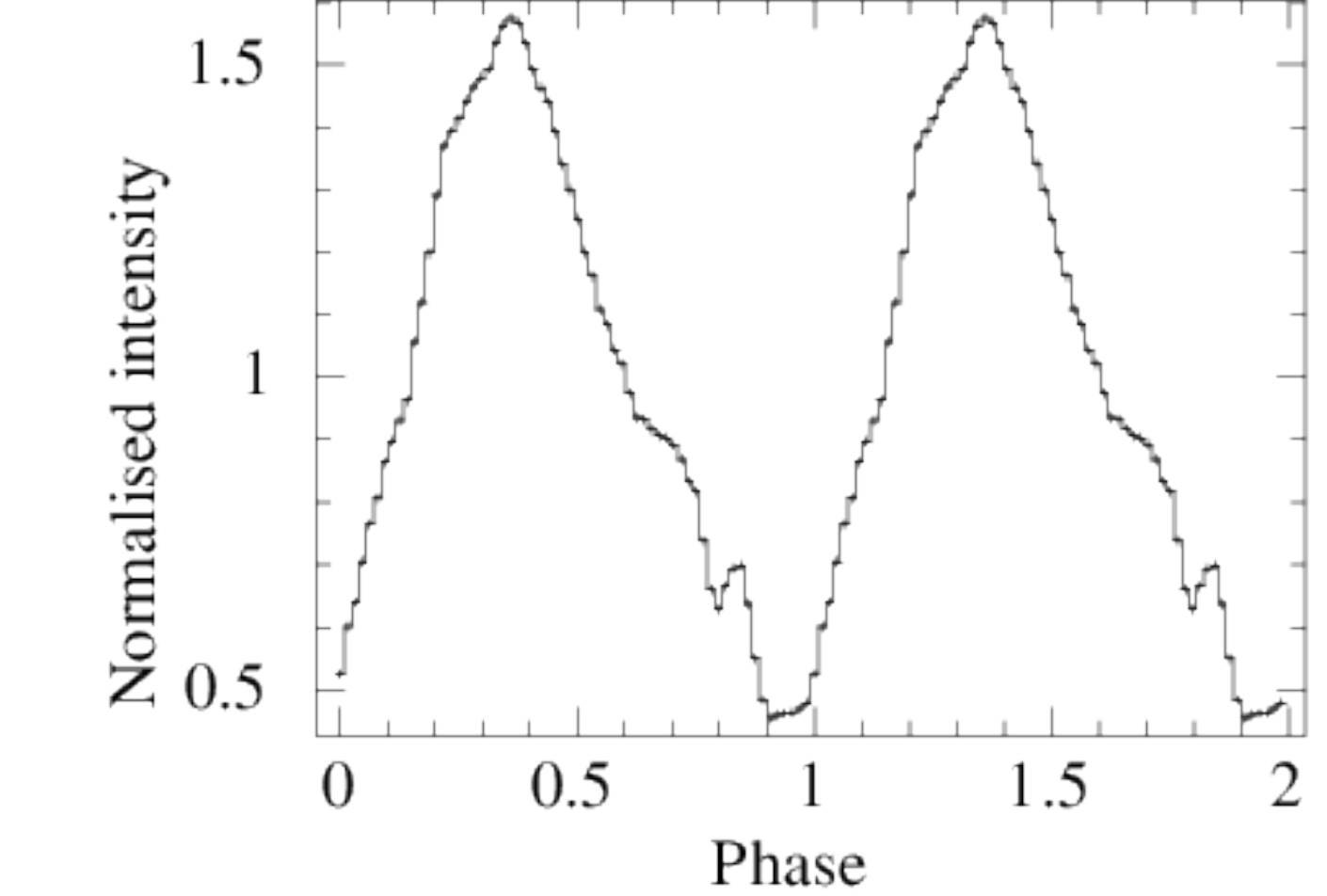}
    \caption{The average pulse profile of \ig in the 3.0--80.0~keV energy band for LXP20 plotted with 64 phase bins per period.}
    \label{Fig:efold}
\end{figure}

To determine the spin period accurately we used the pulse folding and $\chi^{2}$-maximization method using the FTOOLS task \code{efsearch}. We searched for 1000 trial periods around a central pulse period of 12.48~s with a resolution of 0.1 $\mu$s.  A maximum $\chi^2$, indicating the best period, was obtained for a pulse period value 12.485065$\pm$0.000015~s. The quoted uncertainty in the pulse period was determined by generating a 1000 gaussian randomized light curve realizations and carrying out period searches on all of the sample. This was achieved using the method specified in \citet{Boldin2013}. The orbital parameters of this source are not known except its orbital period. Therefore a period search was done on the light curve which is only barycenter corrected. The period value might still not be accurate due to the orbital motion of the source. We have folded the total light curve with the best period obtained from the period search and used MJD 58784.99998 as the reference epoch to create a pulse profile as shown in Figure \ref{Fig:efold}. The pulse profile shows a very broad single peak which is spread across the whole phase range. There is an indication of the presence of another component on the right side of this peak. \\

\begin{figure}
    \centering
     \includegraphics[scale=0.5,angle=0,trim={4cm 0cm 1cm 1cm},clip]{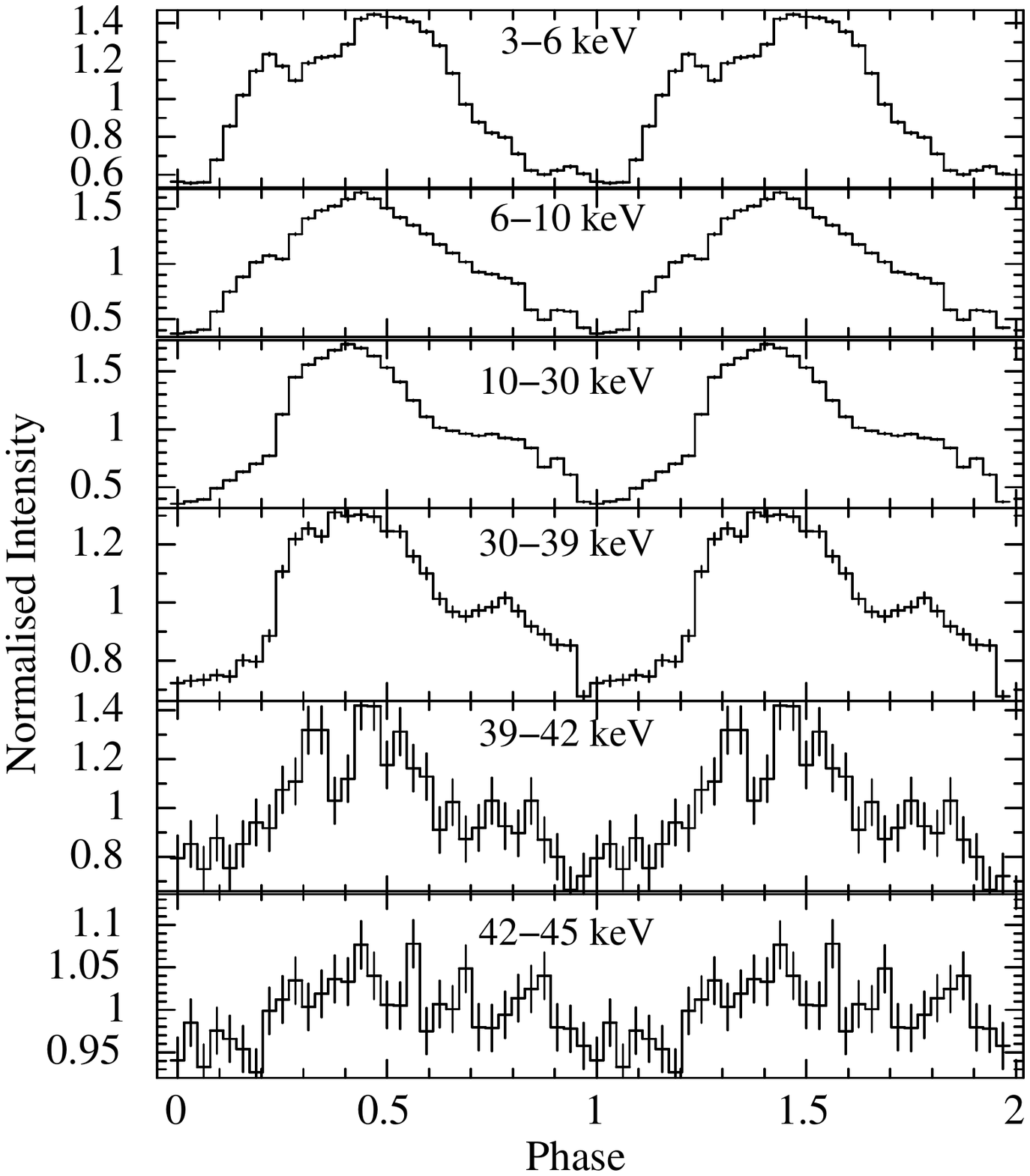}
    \caption{The energy resolved pulse profiles of \ig in 6 different energy bands from the LXP20 detector. These are the energy bands in which pulsations were detected. }
    \label{fig:ene-res-pp}
\end{figure}

\begin{figure}
    \centering
    \includegraphics[scale=0.45]{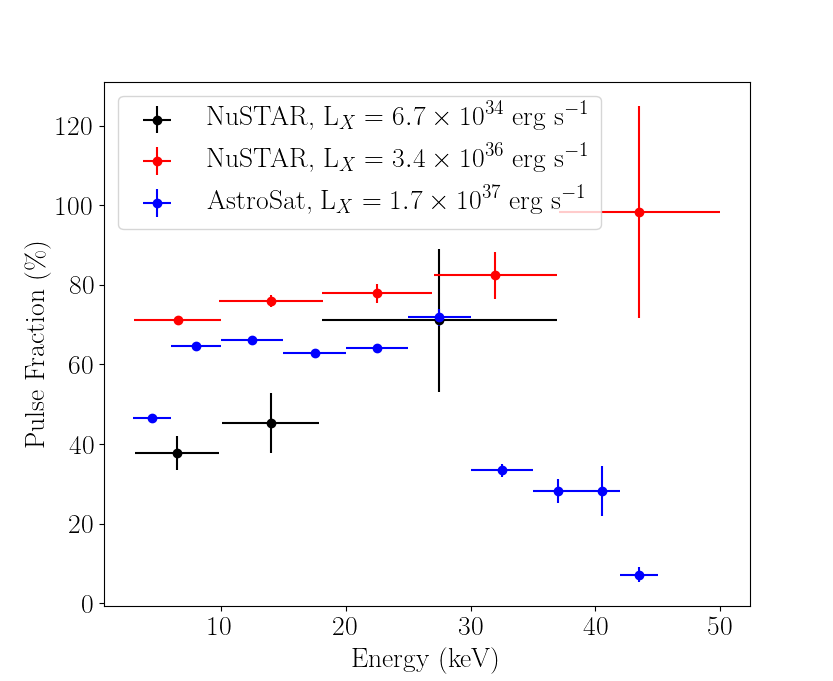}
    \caption{ Variation of pulse fraction is shown for different energy bands. Mid point of energy band is taken as reference for the whole energy band.}
    \label{fig:pff}
\end{figure}

The count statistics in the full energy range data allowed us to create pulse profiles in several energy bands. We extracted light curves in the following bands: 3--6~keV, 6--10~keV, 10--30~keV, 30--39~keV, 39--42~keV, 42--45~keV, 45--65~keV and 65--80~keV. Two energy bands (39--42~keV and 42--45~keV) are specifically chosen around the center of the cyclotron line that was seen in the spectral analysis (see section \ref{sec:spec}). We folded the light curves at the newly estimated spin period and obtained corresponding pulse profiles. The energy resolved pulse profiles are shown in Figure \ref{fig:ene-res-pp}. We detected pulsations upto 45~keV. In the last two energy bands (45--65~keV and 65--80~keV) pulsations are not detected due to poor statistics in the data. The pulse profile has a single peak in the soft X-ray bands. In the 3--6~keV band, the pulse profile has an additional feature to the left of the peak. This feature disappears in the 6--10~keV band. A different component appears on right side in 10--30~keV band and becomes more prominent in subsequent energy bands. We compute the Pulse Fraction (PF) from the pulse profiles using the prescription: \code{$\frac{\rm I_{\rm max}-\rm I_{\rm min}}{\rm I_{\rm max}+\rm I_{\rm min}}$}. The PF shows an interesting trend as a function of energy. The PF first increases from 3--6~keV up to 10--30~keV band and then decreases in the subsequent energy bands as shown in Figure \ref{fig:pff}. We observe this trend in the PF when the source luminosity was $\sim$1.6$\times10^{37}$~erg~s$^{-1}$ (see Section \ref{sec:spec} for spectral analysis and flux estimation details). This trend is different from the one reported by \nus observations that were carried out at two different X-ray luminosities $L_{\rm X}=6.7\times10^{34}$, and $3.4\times10^{36}$~erg~s$^{-1}$, where the PF was found to be increasing with energy in both cases. However, we note that the PF behavior is consistent with the reports of \citet{Roy2017} derived using RXTE data.

\subsection{Spectroscopy} \label{sec:spec}

 We further carried out a joint spectral analysis using the SXT and the LAXPC instruments. The spectral products for the LXP20 detector, which include the source and background spectra in the 5.0--50.0~keV energy range and the response files were extracted for all the layers using the Format A pipeline as specified in Section \ref{sec:obs}. We restrict our spectral range to 50.0~keV beyond which the spectrum has very limited S/N ratio. We also ignore the energy channels below 5.0~keV since the LAXPC response is not fully understood at those energies. The 0.3--7.0~keV SXT spectrum was extracted from the LEVEL 2 event file using the \code{Xselect} tool. For \sxt spectral fitting, we used the background spectrum and Response Matrix File (RMF) distributed by TIFR-POC\footnote{\url{https://www.tifr.res.in/~astrosat_sxt}} and the  \code{sxt$\_$ARFModule} tool to generate the corrected Auxilary Response File (ARF). We then imported all the spectral products into \code{XSPEC} v.12.10.0 for fitting. We included the recommended 3\% systematics when fitting the joint LAXPC--SXT spectra, in order to account for the uncertainty in the spectral responses. \\

\begin{figure}
    \centering
    
    \includegraphics[scale=0.34,angle=-90,trim={0cm 1cm 0cm 0cm}]{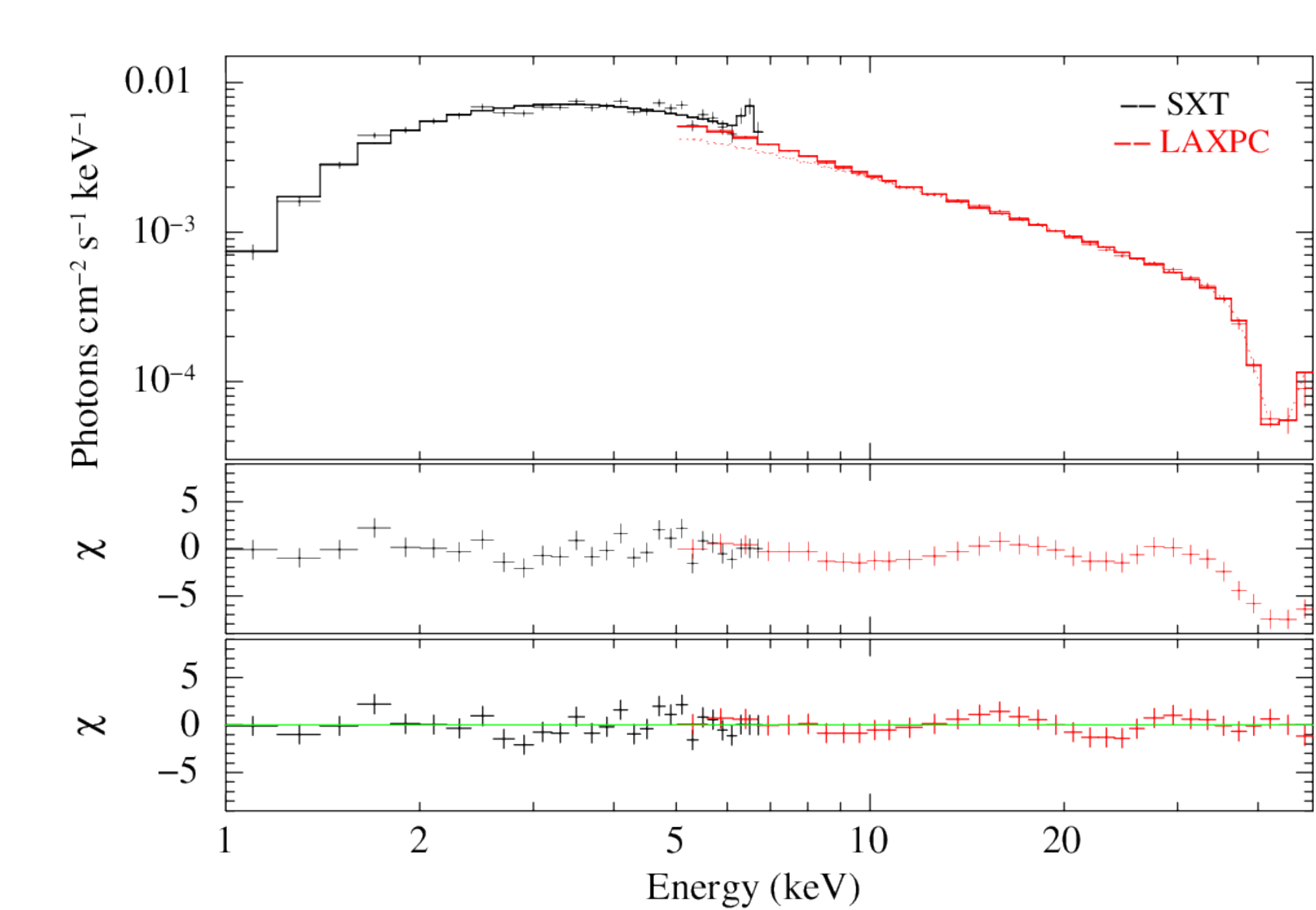}

\caption{The unfolded LXP20 spectrum and fit residuals are shown for the joint fit using SXT and LAXPC obtained for Model 1 `\code{nthcomp}'. The total model fit is shown as a solid black (SXT) and red (LAXPC) lines. The middle panel shows the residuals with no CRSF component added, while the lower most panel shows the residuals when the CRSF line has been accounted for.}

    \label{fig:spectrum}
\end{figure}

\begin{table*}
    \centering
    \caption{Best fit spectral parameters for the joint SXT and LAXPC spectral fit.}
    \begin{tabular}{|c|c|c|c|c|}
    \hline
    &&&& \\
      Model & Parameters  & Model 1 (nthComp)  & Model 2 (PL$\times$highEcut) & Model 3 (cutoff PL)  \\
      &&&& \\
      \hline
     &&&& \\
      \textit{const} &  LAXPC  & 1.0 (fixed) & 1.0 (fixed) &  1.0 (fixed)\\
      & SXT & 1.14  & 1.13 & 1.13\\
      &&&& \\
      \textit{TBabs} & nH ($\times$10$^{22}$~cm$^{-2}$) & 1.65$\pm$0.2 & 2.1$\pm$0.3 & 2.02$\pm$0.4\\
      &&&& \\
      \textit{bbodyrad} & kT (keV) & 1.41$\pm$0.1 & 1.68$\pm$0.1 & 1.57$\pm$0.1\\
      &  norm$^a$ & 1.4$\pm$0.8 & 2.1$\pm$0.6 & 2.24$\pm$0.3\\
      &&&&\\
      \textit{nthcomp} & Gamma & 1.5$\pm$0.1 & &\\
      & kT$_e$ (keV) & 11.2$\pm$2.1  & &\\
      & kT$_{BB}$ (keV) & 1.41$\pm$0.1  & &\\
      & & (tied to kT$_{BB}$) & & \\
      &&&& \\
      \textit{power law} & index ($\alpha$) & & 0.55$\pm$0.1 & \\
      & norm$^b$  & &  0.01$\pm$0.003 & \\
      &&&& \\
      \textit{highEcut} & cutoff E (keV)  & & 14.8$\pm$1.4 &\\
      & fold E (keV) &  & 24.62$^{+9.6}_{-4.1}$ & \\
      &&&& \\
      \textit{cutoffpl} & index ($\alpha$) & & & 0.26$\pm$0.01\\
      & highEcut & & & 18.4$\pm$3.8 \\
      & norm$^b$  & & & 0.01 \\
      &&&& \\
      \textit{gabs}  & E$_{cycl}$ (keV) & 42.7$\pm$0.9 & 45.8$\pm$3.1 & 42.9$\pm$0.8\\
      & $\sigma_{cycl}$ (keV) & 2.72$\pm$1.1 &  4.5$\pm$2.0 & 2.9$\pm$0.8 \\
       & optical depth ($\tau$) & 1.8$\pm$0.1 &1.9$\pm$0.9 & 1.7$\pm$0.5 \\
      &&&& \\
      
      \textit{gaus} & E$_{Fe}$ (keV) & 6.40$\pm$0.1  &  6.42$\pm$0.1 & 6.40$\pm$0.1\\
      & $\sigma$ (keV) & 0.01 (fixed) & 0.01 (fixed) & 0.01 (fixed)\\
      & Eq width (eV) & 128.8 & 130.1 & 136.1 \\
      &&&& \\
      
     \hline
     &&&& \\
    Reduced $\chi^2$ / dof & & 1.1/50&  1.01/50 & 1.03/53\\
      &&&& \\
      \hline
    \end{tabular}
    
    \flushleft $^a$ The \code{blackbody} norm is a dimensionless parameter defined as $\frac{R_{\rm km}^{2}}{D_{\rm 10}^2}$, where $R_{\rm km}$ is the source radius in km and $D_{\rm 10}$ is the distance to the source in units of 10~kpc.
    \flushleft $^b$ photons~keV$^{-1}$~cm$^{-2}$~s$^{-1}$ at 1~keV
    \label{tab:spec}
\end{table*}

The 5.0--50.0~keV LXP20 spectrum and the 1.0--7.0~keV SXT spectrum were jointly fit by introducing a relative normalization term in the form of a constant factor. The constant factor was fixed at 1.0 for LAXPC and was left free to vary for the SXT spectrum. We also applied a gain correction for the SXT pipeline by using the \code{gain$\_$fit} command. We fixed the gain slope to 1.0 and allowed the gain offset value to vary. To describe the spectral contiuuum,  we compared three different models: 1) The thermal comptonization model `\code{nthcomp}' \citep{Zd1996,Z1999} and 2) an absorbed power law with a high energy cutoff, `\code{highEcut}' and 3) a \code{cutoffpl} model. The `\code{nthcomp}' model describes comptonization of soft photons in a hot plasma. It provides a better description of the continuum shape at lower energies compared to a power law by incorporating a low energy rollover such that the scattered spectrum has a fewer photons at energies below the typical input seed photon energies\footnote{https://heasarc.gsfc.nasa.gov/xanadu/xspec/manual/node205.html}. We then compared the spectral fit parameters of this model with the two other phenomenological models, i.e.,  `\code{highEcut}' and `\code{cutoffPL}'. We also note that the more commonly adopted physical model `\code{compTT}' was resulting in highly unphysical fit parameters and was therefore ignored for this analysis. The ISM absorption model \code{TBabs} was used to account for the line-of-sight neutral hydrogen (nH) column density. We adopted the updated photoionization cross sections and the Wilms abundance model \citep{Wilms2000}. All three continuum models showed significant soft X-ray residuals that prompted us to include a blackbody component, which significantly improved the fit. We assume that the same blackbody photons contribute as seed photons for the comptonization process, and so we tie the two temperatures for the fitting. Blackbody temperatures of the order of $\sim$1.5 keV, as seen for this pulsar in this work, were recently reported in a similar transient pulsar system, Swift J0243.6+6124 \citep{GauravJaisawal-2017}. We note that all three spectral models indicate a consistent blackbody temperature with an emission size of $\sim$1.3--1.5~km. 

The LAXPC spectrum using all three models also showed a broad absorption feature near $\sim$42~keV, which we interpret as the CRSF. This line is fit using a Gaussian absorption (\code{gabs}) model. The addition of the \code{gabs} model improved the reduced $\chi^2$ to 1.1 (with a $\Delta\chi^{2}\sim$52.4 for 50 dof). 
The CRSF line energy is poorly constrained by the highEcut model with a larger uncertainty (45.8$\pm$3.1~keV), compared to the other two models. To compute the significance for multiplicative components like the CRSF line, we adopt the F-test routine in IDL package, \code{MPFTEST\footnote{https://pages.physics.wisc.edu/~craigm/idl/down/mpftest.pro}} \citep{DeCesar2013} and compute an F-test probability. We estimate the significance of the CRSF line to be 5.2$\sigma$, 6.0$\sigma$ and 5.8$\sigma$, for Models 1, 2 \& 3, respectively. The SXT spectrum was dominated by continuum emission, with a narrow excess near $\sim$6.4~keV. We added a Gaussian emission line with a fixed width of 0.01~keV to model this line for all three spectral model fits. The Fe K-$\alpha$ line is detected at $\sim$6.4~keV and is independent of the continuum spectral model. Details of the best fit parameters are shown in Table \ref{tab:spec}. We obtain a positive SXT gain offset value of 30.2~eV, 36.1~eV and 30.5 eV, for the three models, respectively. 

In order to statistically examine the CRSF line, we carried out a correlation study of the CRSF line parameters with different continuum parameters. The CRSF centroid line energy at $\sim$42~keV remained largely uncorrelated with the compton continuum parameters. We also carried out an F-test to compare the spectral models. We find that Model 2 (\code{`highEcut'}) is a statistically more preferred phenomenological model (with a p-value$\sim$0.001) compared to the `\code{cutoffPL}' model. However, we find that the physical model `\code{nthcomp}' is statistically indistinguishable from the two phenomenological models using the current data.

\section{Discussion}

 \begin{table*}
    \centering
    \begin{tabular}{|c|c|c|c|c|c|}
    
    \hline
      Sr  & Source & Nature & E$_{cycl}$  & QPOs  & Ref.  \\
      No. & & & (keV) & $\nu_{qpo} $ & \\
      \hline
      &&&&&\\
         1 & 4U 0115+63 & Transient &12,24,36, 48,62 & 1-2 mHz, 27-46 mHz & 1, 2 \\
         2 & V 0332+53 & Transient &28 & 0.22 Hz, 0.05 Hz & 3,4\\
         3 & A 0535+26 & Transient &50 & 27 - 72 mHz & 5, 6 \\
         4 & 1A 1118-61 & Transient &55, 110? & 80 mHz, 70-90 mHz & 7, 8 \\
         5 & Swift 1626.6-5156 & Transient &10,18 & 1 Hz & 9 \\
         6 & Cen X-3 & Persistent & 28 & 40-90 mHz &  10\\
         7 & GX 304-1 & Transient & 54 & 0.125 Hz & 11 \& 12\\
        8 & 4U 1901+03 & Transient & 30 & 0.135 Hz  & 
        13 \& 14\\
        9 & 4U 1626-67 & Persistent & 37 & 1 mHz, 48 mHz & 15, 16 \& 17\\
         10 & IGR J19294+1816 & Transient & 42 &  3.5 mHz &  18 and \textit{\textbf{this work}} \\
         &&&&&\\
         \hline
         
    \end{tabular}
    \color{black}
    \caption{A list of all the transient and persistent X-ray binary sources exhibiting a \textit{CRSF} as well as a QPO. Numbered references to the QPO and CRSF publications are indicated below. 1.  \citet{Jayashree2019} 2. \citet{Heindl1999} 3. \citet{Qu_2005} 4. \citet{caballerogarcia2015} 5. \citet{Finger1996} 6. \citet{CameroArranz2012} 7. \citet{Jincy2011} 8. \citet{Maitra2012} 9. \citet{Reig2008} 10. \citet{Raichur2008} 11. \citet{Devasia2011} 12. \citet{Yamamoto2011} 13. \citet{Marykutty2011} 14. \citet{Beri2020} 15. \citet{Orlandini1998}  16. \citet{Raman2016} 17. \citet{Chakrabarty2001} 18. \citet{Tsygankov2019}.}
    \label{tab:cycl-qpo}
\end{table*}

We report the results from a timing and spectral study of the HMXB pulsar IGR 19294+1816 using \asr observations that were carried out during the October 2019 outburst. The observations fall on the declining phase of the outburst. We measure the pulsar's spin period to be 12.485065$\pm$0.000015~s. We also report the detection of a 0.032~Hz QPO from the timing analysis of LAXPC light curves. From our spectral analysis results we find the cyclotron line energy to be centered at 42.7$\pm$0.9~keV (consistent with \nus report, \citealt{Tsygankov2019}) and estimate the pulsar's magnetic field to be 4.6$\times$10$^{12}$~G. From the 0.7--50~keV model flux obtained using our spectroscopy results and assuming a source distance of 11~kpc, we find that during its declining outburst phase, IGR 19294+1816 had a source luminosity of 1.6$\times$10$^{37}$~ergs s$^{-1}$.

\subsection{Cyclotron line \& Quasi Periodic Oscillation}

Out of 36 X-ray binary sources with confirmed detections of CRSF \citep{Staubert2019}, only 9 sources exhibit simultaneous cyclotron line absorption along with a QPO.This becomes crucial since we can now obtain two independent measures of the magnetic field and probe the magnetospheric interactions using combined timing and spectral properties. A list of such sources are given in Table \ref{tab:cycl-qpo}. Results from our work makes IGR 19294+1816 the 10th such source in this list. 

\begin{figure}
    \centering
    
    \includegraphics[scale=0.42,angle=0,trim={0cm 0.0cm 0 0.1cm},clip]{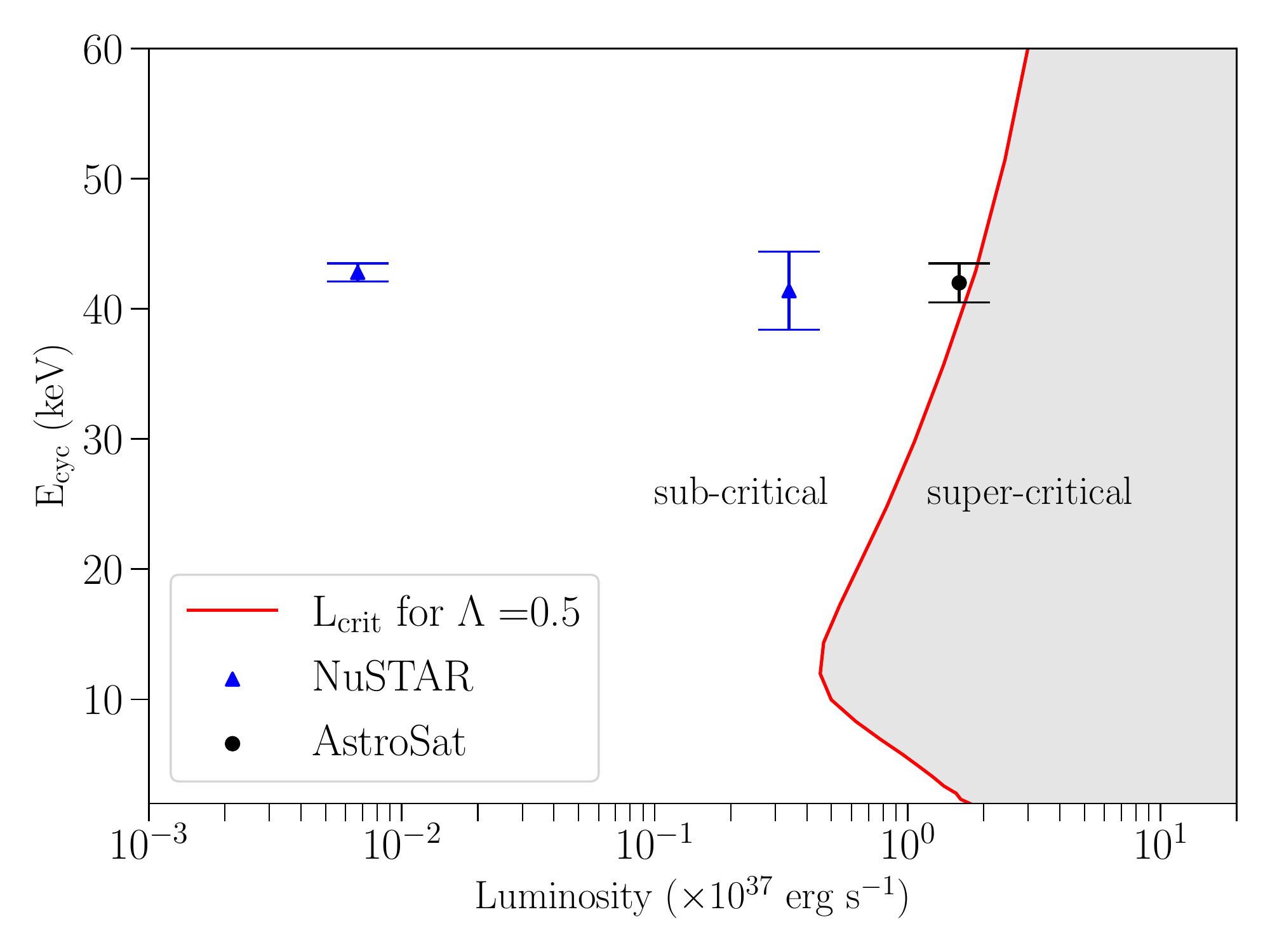}
    \caption{The CRSF line energy measured by \nus and \asr for IGR 19294+1816 are shown as function of luminosity. The red curve indicates the critical luminosity model function (for an accretion disk parameter of $\Lambda=0.5$) that demarcates the sub and super critical accretion regimes \citep{Mushtukov2015}.}
    \label{fig:cyc-lum}
\end{figure}

 In our work, we detect the CRSF feature at 42~keV, which indicates (for a z of $\sim$0.3 and for M$_{NS}$ in the range 1.4--2.0 M$_{\odot}$: \citealt{Cottam2002}) a magnetic field strength of 4.6$\times10^{12}$~G. Our measurement of E$_{\rm CRSF}$ agrees with the \nus studies of the source \citep{Tsygankov2019}. It is also interesting to note that the centroid energy of the cyclotron line has remained steady between the different luminosity states. Particularly, \citet{Tsygankov2019} have conducted their spectral analysis over two luminosity states  (L$_{\rm X}\sim$6.4$\times10^{34}$~ergs~s$^{-1}$ and L$_{\rm X}\sim$3.4$\times10^{36}$~ergs~s$^{-1}$) and our work probes a slightly higher luminosity (i.e. L$_X\sim$1.6$\times10^{37}$ ergs~s$^{-1}$) during which the E$_{\rm CRSF}$ has remained fairly steady at $\sim$42~keV (see Figure \ref{fig:cyc-lum}). 

Interestingly, there is strong observational evidence for the bimodal dependence of the CRSF line energy with X-ray luminosity (L$_{\rm X}$) \citep{Becker2012,Mushtukov2015, Staubert2019}. Both positive and negative correlations between the CRSF line energy and L$_{\rm X}$ have been observed in various HMXB pulsars and the luminosity where the correlation changes from positive to negative has therefore been associated with the critical luminosity. \citet{Becker2012} also provide a theoretical framework that explains the beam pattern in both the regimes - above L$_{\rm crit}$, radiation dominated shocks are responsible for the generation of a fan beam; below L$_{\rm crit}$, Coulomb interactions dominate the flow decelaration process and radiation escapes parallel to the field to produce a pencil beam pattern. 

In this context we discuss different works addressing the computation of the critical luminosity achieved in pulsar accretion columns. \citet{Becker2012} derive an effective L$_{Edd}$ that arises specifically for the case of pulsar accretion columns. The overall accretion dynamics is governed by matter being decelerated in response to radiation pressure or coulomb interactions and these two domains are separated by a critical luminosity (L$_{\rm crit}$), which for a canonical NS, is $\sim$10$^{37}$~erg s$^{-1}$. However, more recent calculations by \citet{Mushtukov2015}, which are based on the physical model derived by \citet{Basko1976}, provide an accurate estimate of critical luminosity by accounting for the resonances in the Compton scattering cross section, polarization as well as the accretion flow configuration. They show that the L$_{\rm crit}$ is not a monotonic function of the magnetic field and has a more complicated behavior. Their theoretical work compares well with observed behavior for some of the sources like V 0332+53, 4U 0115+63, GX 304-1. Sources like A 0535+26 show neither a positive or a negative correlation \citep{Caballero2007,Mushtukov2015}. The CRSF line in IGR 19294+1816 has so far been measured using \nus and \asr. Figure \ref{fig:cyc-lum} shows the CRSF line energy measurements as a function of luminosity along with the critical luminosity model curve (shown for an accretion disk parameter of $\Lambda=0.5$, \citealt{Mushtukov2015}). The current \asr measurement lies very close to the L$_{\rm crit}$, in the sub-critical regime. Future broadband X-ray observations in different luminosity states might be useful in understanding the accretion geometry and in turn, the correlation behavior of this pulsar. 

In accretion powered X-ray pulsars, both the transient as well as the persistent population often exhibit QPOs. They are observed for several of the sources during their outbursts (see \citealt{bpaul2011} and references therein). Since HMXBs host highly magnetized NS, their inner disk radii extend to about 1000 km due to which their corresponding keplerian frequencies fall in the mHz regime. We can therefore expect to detect slow mHz QPOs ($\sim$1~mHz--1 Hz) in such systems \citep{Psaltis2006}. This is unlike QPOs observed in LMXBs or black hole X-ray binaries whose QPO frequencies are $\sim$1~Hz--100's of Hz. 

QPOs are traditionally understood to be a result of plasma instabilities that are generated around the magnetospheric boundary. Two very popular models that are often invoked to explain the QPO frequency ($\nu_{qpo}$) are the Keplerian Frequency Model (KFM, \citealt{vanderklis1987}) and the Beat Frequency Model (BFM, \citealt{AlparShaham1985}). In the KFM, the X-rays are generated by inhomogeneities that are modulated at $\nu_{k}$ and so the QPO is generated at the keplerian frequency corresponding to the inner accretion disk. Thus the keplerian frequency is $\nu_{\rm k}=\nu_{\rm qpo}$. However, the KFM is valid under the assumption that the NS is spinning slower than the keplerian frequency corresponding to the radius at which the QPO generating inhomogeneity is located. This is the case for some of the HMXBs like EXO 2030+375, A0535+262 \citep{Finger1996}, etc. Otherwise, a faster rotating NS will give rise to centrifugal forces that will inhibit accretion \citep{Stella1986}. The BFM, on the other hand, assumes that the keplerian frequency is modulated by the rotating magnetic field of the NS. The QPO feature is therefore assumed to arise as a beat phenomenon between the NS spin ($\nu_{\rm spin}$) and the $\nu_{\rm k}$, giving $\nu_{\rm qpo}=\nu_{\rm k}-\nu_{\rm spin}$.  

IGR 19294+1816 has a spin frequency of $\nu_{\rm s}=0.08$~Hz and the measured QPO frequency lies at $\nu_{\rm qpo}\sim 0.034$~Hz. Since the $\nu_{\rm s} > \nu_{\rm qpo}$, the KFM does not apply to this object in this scenario.  In order to examine if the BFM is applicable, we use the X-ray luminosity and the QPO frequency measurements to estimate the NS surface magnetic field as predicted by the BFM. Assuming a circular orbit, we can calculate the radius at which the QPO is generated as,
 \begin{equation}
    r_{\rm qpo}=\bigg(\frac{GM_{\rm NS}}{4\pi^2\nu^2_{\rm k}}\bigg)^{\frac{1}{3}}
\end{equation}
where, the keplerian frequency as predicted by the BFM is, $\nu_{\rm k}=\nu_{\rm qpo}+\nu_{\rm spin}$. The BFM further assumes that the QPO is generated at the magnetospheric radius (r$_{\rm M}$), i.e., r$_{\rm M}=r_{\rm qpo}$. According to \citet{GhoshLamb1979}, r$_{\rm M}$ is given by:

\begin{equation}
    r_M=2.3\times10^8 \times M_{1.4}^{1/7}  R_{6}^{10/7}  L_{37}^{-2/7} B_{2.5B_{12}}^{4/7} cm
\end{equation}
 For a generic NS with mass $\sim$1.4 $\rm M_{\odot}$ and a measured 1.0--50.0~keV flux of 1.2$\times$10$^{-9}$~ergs~s$^{-1}$ cm$^{-2}$, we obtain a magnetic field strength in the range 1.2$\times$10$^{13}$--6.2$\times$10$^{12}$~G, corresponding to a neutron star with radius in the range 10--15~km.

Another model that has been invoked to explain such low frequency mHz QPOs is the magnetic disk precession model \citep{Shirakawa-Lai2002}. The authors demonstrate that the inner region of the accretion disk is subject to magnetic torques that can cause warping and precession of the accretion disk. Under typical conditions present in X-ray pulsars, these magnetic torques can overcome the viscous damping and can enable the instability mode to grow, which may in turn generate mHz QPOs. This model was proposed by \citet{Shirakawa-Lai2002} originally to explain the mHz QPOs in sources like 4U 1626-67, etc. More recently,  \citet{Dugair2013} and \citet{Roy2019} showed that this warped disk model was able to explain the mHz QPOs observed in 4U 0115+63.  

The QPO precessional frequency as specified by \citet{Shirakawa-Lai2002} is given by:
\begin{equation}
    \tau_{prec} = 776\times \alpha^{0.85} \times L_{37}^{-0.71} \rm s
\end{equation}
where $\alpha$ is the accretion disk viscosity parameter (usually $<$1), and L$_{37}$ is the X-ray luminosity in units of 10$^{37}$~erg/s. Assuming an $\alpha$=0.02 (this choice comes from the model fits carried out by \citealt{Jayashree2019} for the source 4U0115+63), we obtain a predicted $\nu_{\rm qpo}\sim$3.6~mHz, which is a factor of 10 smaller than our detected 32~mHz QPO.

\subsection{Broadening of the pulse peak}
The base of the pulse peak in the power density spectrum from our analysis of IGR 19294+1816 exhibits a broadened shape very similar to the one previously observed by \citet{Rodrig2009}. These wing like features have also been previously reported in several other HMXB sources, for example, Cen X-3 \citep{HarshaPaul2008}, 4U 1626-67 \citep{JainPaulDutta2010}, GX 304-1 \citep{Jincy2011}, 4U 1901+03 \citep{Marykutty2011} and Her X-1 \citep{MoonEiken2001}. In order to explain wings observed in HMXB sources, \citet{LazztiStella1997} and \citet{Burderi1993} have shown that the periodic and aperiodic variability components of the power spectrum are not to be considered independent of each other, as has been traditionally assumed. They show that these wings arise as a result of the coupling between the coherent periodic pulse variability and the aperiodic red noise component. We also note that the QPO feature is observed simultaneously alongside the broadened peak unlike in some other sources where the pulse peak narrows down once the QPO starts appearing (see for example, \citealt{Marykutty2011}). A clear understanding of such anomalies requires a systematic study using of transient pulsars across multiple observations. 

\subsection{Outbursts in transient Be X-ray binaries}

IGR 19294+1816 has been repeatedly studied during each of its previous outbursts: the very first 2009 outburst using \textit{INTEGRAL} \citep{Atel2}, the second outburst in October 2010  using \textit{INTEGRAL} \citep{Bozzo2011} the combined  2009 \& 2010 outbursts using RXTE \citep{Roy2017}, and the further outbursts in 2017-2018 using \textit{Swift} and \nus \citep{Harrison2013, Tsygankov2019}. Our study of this pulsar during its most recent 2019 outburst using \asr has additionally strengthened our knowledge about some of the timing and spectral characteristics of this source. 

The detection of a pulse peak and its higher harmonics have been reported in all previous outbursts including in this current work. Notably, the higher harmonics decrease in strength as the outburst progresses from its highest to its lowest flux levels as seen from RXTE observations \citep{Roy2017}. Inhomogeneities in the NS accretion disk also get prominently detected in the form of mHz QPOs in a number of HMXB pulsars during their periastron passage outbursts (for example, 4U 0115+63: \citealt{Dugair2013}, V 0332+53:\citealt{Caballero-Garc-2016}, etc.) and IGR 19294+1816 is no exception to this. The strong presence of the mHz QPO (with an increasing rms power with energy) for IGR 19294+1816 at such a high luminosity, possibly indicates that the inner regions surrounding the magnetosphere become most visible during the highest flux states. Interestingly, this is in stark contrast with the behavior observed for another HMXB transient, V0332+53 \citep{Caballero-Garc-2016}, where the mHz QPO is most visible during the lowest flux states (and the rms power of the QPO decreases at higher energies, see \citealt{Qu2005}). 

Another interesting outburst characteristic is the variation of the pulse fraction. RXTE observations \citep{Roy2017} and LAXPC observations (this work) tend to show an increasing pulse fraction all the way up to $\sim$30 keV, as expected in most accretion powered X-ray pulsars. However, beyond 30 keV, the PF drops dramatically. This is a peculiar deviation from the trend reported using \nus \citep{Tsygankov2019} for this source. Although the current LAXPC observations have captured the source during a much higher luminosity compared to the \nus observations, we note that the PF is lower.   

Unlike the other transient Be X-ray pulsars that exhibit luminosity dependent pulse profiles (for example, A0535+262, 1A 1118-61, etc. see \citealt{bpaul2011} and references therein), IGR 19294+1816 has not demonstrated any detectable changes in the pulse profiles as a function of different luminosity states as also pointed out by \citet{Tsygankov2019}. It has also not exhibited any luminosity state dependence for the CRSF parameters within the last decade. It does however show a variation of pulse profiles as a function of energy. Most pulsars exhibit a dependence of profile shape with energy as well as time \citep{WhiteSwankHolt1983}. At different source luminosities, the appearance of complex sub structures in the pulse profiles has been reported in some sources like GX 304-1 \cite{Devasia2011}. The changes in the pulse profile shape seen for IGR 19294+1816 and similar pulsars can be attributed to the variable geometry and physical processes that occur near the NS surface and within the accretion column.

\section{Conclusions}
We have analysed the outburst characteristics of the transient X-ray pulsar IGR 19294+1816 using the LAXPC and SXT instruments on board \asr during the falling phase of its 2019 outburst. We have detected a low frequency QPO at 0.032~Hz; a similar feature (at $\sim$0.035~Hz) was tentatively reported earlier by \citet{Rodrig2009} using RXTE. We have also confirmed the presence of a strong cyclotron feature at 42~keV and a fairly weak Fe emission line at 6.4~keV using the joint SXT and LAXPC broadband spectra. Our studies indicate that some spectro-timing features for IGR 19294+1816 are consistent with other transient Be X-ray binaries and some properties are not. We encourage future broadband observations with good spectral resolution that can sample different accretion regimes in order to test various E$_{\rm CRSF}$/L$_{\rm X}$ correlations for this object.
Obtaining stronger binary parameter constraints also will help understand this class of transient Be X-ray binary sources better.  \\

\textit{Acknowledgements} : This publication is based on the results obtained from the \asr mission of the Indian Space Research Organisation (ISRO), archived at the Indian Space Science Data Centre (ISSDC). We thank members of LAXPC instrument team for their contribution to the development of the LAXPC instrument. We also thank the LAXPC \& SXT POC at TIFR for verifying and releasing the data via the ISSDC data archive and providing the necessary software tools. The authors are grateful to Alexander Mushtukov for providing us with the latest critical luminosity model curve that have been used in this study.\\

\noindent\textbf{Data availability statement:}\\
The data underlying this article are available in the AstroSat data archive: https://astrobrowse.issdc.gov.in/astro$_:$\\archive/archive/Home.jsp \\

\bibliography{bibtex}{}
\bibliographystyle{mn2e}

\end{document}